# An Empirical Study of End-User Behaviour in Spreadsheet Error Detection & Correction

Brian Bishop, Dr. Kevin McDaid
Dundalk Institute of Technology,
Dundalk, Ireland
brian.bishop@dkit.ie, kevin.mcdaid@dkit.ie

**ABSTRACT**

*Very little is known about the process by which end-user developers detect and correct spreadsheet errors. Any research pertaining to the development of spreadsheet testing methodologies or auditing tools would benefit from information on how end-users perform the debugging process in practice. Thirteen industry-based professionals and thirty-four accounting & finance students took part in a current ongoing experiment designed to record and analyse end-user behaviour in spreadsheet error detection and correction. Professionals significantly outperformed students in correcting certain error types. Time-based cell activity analysis showed that a strong correlation exists between the percentage of cells inspected and the number of errors corrected. The cell activity data was gathered through a purpose written VBA Excel plug-in that records the time and detail of all cell selection and cell change actions of individuals.*

## 1. INTRODUCTION

The ubiquity of spreadsheet programs within all levels of management in the business world indicates that important decisions are likely to be made based on the results of these, mainly end-user developed, programs. The financial sector is particularly dependent on spreadsheets [Croll, 2005]. Unfortunately, the quality and reliability of spreadsheets is known to be poor following empirical and anecdotal evidence collected on the subject [Panko, 1998], [Rajalingham et al, 2000] and [Chadwick, 2004]. From the experience of one consulting firm, Coopers and Lybrand in England, 90% of spreadsheets with over 150 rows of data were found to contain one or more faults [Panko, 1998], and due to the nature of spreadsheets, when failures do occur, the results can be quite significant. For example, sudden budget cuts were necessary at the University of Toledo after an erroneous spreadsheet formula inflated projected annual revenue by $2.4 million [Fisher et al, 2006].

Many spreadsheet auditing tools have been developed and are widely available, but to develop auditing tools that compliment end-users natural auditing and debugging behaviour, research into this behaviour needs to be conducted. To date, we have found only one study that addresses end-user behaviour/processes in the inspection and debugging of spreadsheets, [Chen & Chan, 2000]. The study was somewhat limited as cognitive processes were captured using video taping and a thinking-aloud protocol from four participants without spreadsheet and accounting expertise. To this end, we undertook to investigate and unintrusively record the behaviour of industry-based professionals and students during the spreadsheet debugging process. Thirteen industry-based spreadsheet developers and 34 accounting and finance students took part in the experiment.

The layout of the paper is as follows. Section 2 introduces the topic of spreadsheet error detection and correction and compares the activity of spreadsheet inspection and debugging with that of imperative programming language verification, validation and debugging. Section 3 details our research goals and experiment methodology. In Section 4 initial findings of the experiment are presented. A conclusion and proposed future research are detailed in Section 5.

## 2. SPREADSHEET ERROR DETECTION AND CORRECTION

Very little research has been conducted on the error detection process for spreadsheets. The emphasis of the small amount of spreadsheet research available has been on the prevention of spreadsheet errors through spreadsheet design and testing methodologies. The notable exceptions to this are [Chen & Chan, 2000] which is mentioned in the previous section, [Galletta et al, 1993], [Galletta et al, 1996], [Panko, 1999] and [Howe & Simpkin, 2006] in which studies on error-finding performance, the effect of spreadsheet presentation in error detection, applying code inspection to spreadsheet testing and the factors affecting the ability to detect spreadsheet errors were undertaken respectively. Importantly, none of these papers, unlike our work, deal with the cell-by-cell processes by which, or the order in which, these errors are found and corrected. In [Galletta et al, 1996], the author concludes that an increased understanding of the error-finding process could help avert some of the well publicised spreadsheet errors.

Authors have in the past looked to the traditional software development domain for methods and tools that could yield spreadsheet process improvement. Examples include the application of code inspection to spreadsheet testing [Panko, 1999], software visualisation applied to spreadsheets for fault localisation [Ruthruff et al, 2003] and using test driven development, an eXtreme Programming technique, for developing spreadsheets [Rust et al, 2006]. The application of traditional software verification and validation (V&V) and debugging research to the spreadsheet paradigm would seem like a natural course of action for spreadsheet error detection and correction research, but error detection and debugging in spreadsheets is a combination of static and dynamic V&V, and debugging associated with traditional programming languages. A comparison between spreadsheets and traditional programming languages and the ramifications of the differences with regard to testing techniques and terminology are discussed briefly in the next section.

### 2.1 Software Verification and Validation: Spreadsheets Vs Traditional Programming Languages

In traditional software development, verification and validation (V&V) are processes used to determine if the software is being built correctly and if it the correct software is being built, respectively [Jenkins et al, 1998]. V&V is concerned with establishing the existence of defects in a system, as distinct from debugging, which is concerned with locating and repairing these defects once their existence is established [Sommerville, 2004].

Software verification is composed of both static and dynamic verification. Static verification involves the inspection of code and other development artifacts. It is static in nature in that the system is not exercised in the examination of the code or documents. Dynamic verification involves program testing, concerned with exercising and observing program behaviour [Sommerville, 2004]. They are two distinct processes. In the spreadsheet paradigm, code inspection (individual phase) and the debugging process invariably become amalgamated. In the spreadsheet debugging process, where the sought error is found and corrected, the spreadsheet updates all cells. In this way the implication of the change can been seen immediately. This is similar to the dynamic verification process of software testing. What can be taken from this is that static verification, dynamic verification, and debugging in traditional software development are three distinct processes, whereas spreadsheet debugging involves the integration of these into a single process. Further differences between spreadsheet and imperative programming paradigms, and the ramifications these differences may have for spreadsheet testing methodologies have been addressed in [Rothermel et al, 2001].

The terms spreadsheet auditing, debugging, inspection, testing, error finding etc. are sometimes used interchangeably in spreadsheet literature. For the purposes of this paper, spreadsheet error detection refers to the inspection by an individual of the spreadsheet code and the discovery of any errors, error correction refers to the successful correction of discovered errors, and spreadsheet debugging refers to the combination of these processes.

## 3. ANALYSING END-USER BEHAVIOUR: RESEARCH GOALS AND METHODOLOGY

### 3.1 Research Questions

Some questions we seek to answer are as follows:

- Will industry-based spreadsheet developers outperform students in detecting and correcting certain types of errors?
- Is there a correlation between the number of cells inspected and error detection and correction performance?
- Can common patterns of spreadsheet debugging behaviour be identified, and are there particular patterns that are more effective than others?

The research is in its early to mid stages, and many questions and hypothesis have yet to be answered and tested respectively. Initial analysis and some interesting findings from our first experiment are presented in Section 4. The following section details the experiment methodology.

### 3.2 Methodology

**Experimental Spreadsheet Model**

Regarding the detail of the experiment, a spreadsheet model has been developed consisting of three worksheets seeded with errors (see Appendix A). The names and functions of each of the three worksheets are as follows: *Payroll*, compute typical payroll expenses; *Office Expenses*, compute office expenses; *Projections*, perform a 5-year projection of future expenses. Each worksheet has different error characteristics. *Payroll* has data entry, rule violation and formula errors; *Office Expenses* has clerical, data entry and formula errors; *Projections* has mostly formula errors.

Participants were asked to debug the spreadsheet, and each error found was to be corrected directly in the spreadsheet itself. The spreadsheet model was adapted from a previous experiment carried out by Howe & Simpkin [2006], in which 228 students took part in an experiment designed to identify the factors which influence error-detection capability. Among other advantages, using a similar spreadsheet model to the one detailed in [Howe & Simpkin, 2006] allows us to compare results obtained from a large number of students with those of industry-based professionals.

Although other error classification systems exist [Teo & Tan, 1997], [Panko, 1998] , the error classification system from [Howe & Simpkin, 2006] was utilised for this experiment, mainly to allow for detailed comparisons to be made between the error detection results of the 13 professionals and 34 students from our study and the 228 students from the experiment detailed in [Howe & Simpkin, 2006]. The error categories, and number of seeded errors of each category, are as follows. *Clerical/Non material errors* (4), such as spelling errors. *Rule Violations* (4) are cell entries that violate company policy, for example paying an employee overtime when that employee is not eligible for overtime. *Data Entry errors* (8) include negative values, numbers entered as text etc. *Formula Errors* (26), such as inaccurate range references, illogical formulas etc.

**Spreadsheet Cell Activity Tracking Tool**

Crucially, a tool has been developed (in VBA) to record the time and detail of all cell selection and cell change actions of individuals while debugging a spreadsheet. The data recorded is as follows: individual cells selected, cell ranges selected, worksheet selections, individual cells edited & the resulting cell value, cell ranges edited & resulting cell values. Timestamps are recorded for all of these activities (in milliseconds). More complex spreadsheet activities can also be identified from the resulting data log. These include copy and past, undo typing, redo typing and drag-and-fill.

**Sample**

Thirteen industry-based spreadsheet developers took part in the experiment, along with thirty-four second year accounting & finance students. The backgrounds of the professional participants are as follows: Accountants, 8; Financial Analysts, 2; Actuaries, 1; Software Developers, 2; with all the professional participants, including the two software development participants, having an industry based working knowledge of spreadsheet development and use. The need for spreadsheet experiments with industry professionals as opposed to the student population has been voiced by many researchers in this area, including the authors of the aforementioned paper, Howe & Simpkin [2006]. For comparison purposes the experiment was also carried out with 34 second year accounting and finance students.

**Process**

The subjects were given a copy of the experimental spreadsheet along with an instructions page. A short introduction on the instructions page explained the purpose of the task, namely to investigate how effectively spreadsheet users discover and correct errors. Subjects were asked to correct any errors found directly on the spreadsheet itself. The instructions page also contained some rules with regards to the data in the worksheets e.g. only employees with codes B or C are eligible to receive overtime pay.

Both the spreadsheet and the instructions were emailed to each of the industry-based professional subjects after they had been contacted and had agreed to take part. The student subjects were given the opportunity during a single 60 minute class period to participate in the study. No time limit was given to the professional subjects, but as pretests suggested, professional subjects completed the task in an average of 28 minutes, and student subjects in an average 36 minutes. The students knew in advance that a spreadsheet debugging exercise had been arranged, and the general feeling was that participants, both professional and student, approached the task as an interesting challenge. Subjects were told that cell activities were being recorded during the debugging process, and that all individual results would remain confidential. Subjects were given contact details of the authors to request individual results.

## 4. INITIAL FINDINGS

Initial findings from analysis of the data recorded during the experiment and error correction results are detailed in the following sections.

### 4.1 Overall Results

**Error Correction Rates**

Industry-based professional subjects corrected 72% of all seeded errors and student subjects corrected 58% of all seeded errors. The results from Figure 1 show a clear distinction between performances of industry based professionals and students for Rule Violation and Formula errors, with professionals correcting 16% more formula errors than students and 20% more rule violation errors.

| Error Type | No. of Seeded Errors | % Errors Corrected by Professionals | % Errors Corrected by Students | Professionals Compared to Students | [Howe & Simpkin, 2006] Students |
|---|---|---|---|---|---|
| Clerical/Non-Material | 4 | 17% | 11% | + 6% | 66% |
| Rules Violation | 4 | 85% | 65% | + 20% | 60% |
| Data Entry | 8 | 68% | 63% | + 5% | 72% |
| Formula | 26 | 79% | 63% | + 16% | 54% |
| **Total** | | **72%** | **58%** | **+ 14%** | **67%** |

**Figure 1 – Error Correction Results**

**Error Correction Performance: Expert Vs Novice**

The spreadsheet used in this experiment is almost identical (but for six less clerical and five more material errors) to that used by the authors of [Howe & Simpkin, 2006]. This allows for a detailed comparison between results. The mean error detection rate from [Howe & Simpkin, 2006] was 67%, with subjects detecting 66% clerical, 60% rule violation, 72% data entry, and 54% formula errors. Students from both experiments yielded similar results, with the exception of clerical errors. The spreadsheet used by [Howe & Simpkin, 2006] had 10 clerical errors, and students were informed that clerical/spelling errors may be on the spreadsheet; 66% were detected. Students and professionals from our experiment found only 11% and 17% of the 4 clerical errors respectively. They showed little interest in detecting them, and most professionals thought them irrelevant. When comparing the *formula* and *rule-violation* error detection rates of students in [Howe & Simpkin, 2006] and the professional subjects in our experiment, the professional subjects corrected 25% more *formula* errors and 25% more *rule-violation* errors.

Although the overall average error correction results shown in Figure 1 are similar to those of [Howe & Simpkin, 2006], the difference in error correction rates for some of the error categories, particularly *formula errors*, between professionals from our experiment and students from both experiments is quite significant. This finding is contrary to the findings presented in [Galletta et al, 1993], in which spreadsheet experts did not outperform novices in detecting spreadsheet formula errors. The findings from [Galletta et al, 1993] suggest that spreadsheet expertise is not crucial for discovering errors strictly affecting spreadsheet formulas and structure. As the spreadsheet model used in our experiment was not complicated and did not require much, if any, domain knowledge, and given that the participating students in our experiment were themselves accounting & finance students (diminishing further the domain expertise factor), this leads to a conclusion that industry-based professionals with a good working knowledge of spreadsheets i.e. experts, find and correct more formula errors in less time than end-users with little or no industry-based spreadsheet experience i.e. novices.

One important element that may be a contributing factor to the differences in the findings of this experiment and that of [Galletta et al, 1993], with regards to spreadsheet expertise and *formula error* detection and correction performance, is the employment of self-reported measures to establish spreadsheet expertise used in [Galletta et al, 1993]. The professional subjects who took part in the experiment detailed in this paper were known to have a good industry-based working knowledge of spreadsheet use and development prior to their involvement in the experiment.

### Determining Possible Group Code-Inspection Phase Benefits

In this experiment subjects inspected the spreadsheets individually; there was no group code inspection phase of any kind. In an attempt to determine what the average error correction yield might have been if individuals had been placed in groups of three, all combinations of three students were created and the number of separate errors corrected identified. The average performance across all combinations was calculated to represent the likely performance on average of a group of three students. This method assumes that there is no added benefit of working in groups of three beyond the sharing of information. The result of this process suggested that on average 81% of the errors would have been corrected if result pooling in groups of three was performed successfully. In a past study [Panko, 1999] the individual spreadsheet code inspection phase was followed by a group code inspection phase, using groups of three. It was found in that study that the gain from group-work came only from pooling the different errors detected previously by the individuals. The result of 81% is consistent with the 83% yield from group-work in [Panko, 1999].

### 4.2 Cells Inspected Versus Debugging Performance

With the data recorded during the spreadsheet debugging process it is possible to identify any cells that were inspected or edited. It is also possible to determine the number of times each cell was inspected and to determine the time spent inspecting each cell. An important research goal was to determine if there was a correlation between the number of cells inspected and error detection/correction performance. To answer this question, analysis was conducted to identify, for each subject, the number of individual cells inspected or edited during the debugging process. A cell was considered inspected/checked if that cell was selected for a specified minimum time or if the cell value/formula was edited or changed directly, and if the cell was within a specified range of cells. The specified ranges of cells for this analysis were cells that contained formulas/values. Blank cells and column/row headings were not included. For this analysis there were 44 usable results, as time data had not been recorded correctly for one of the students and two of the professionals.

Figure 2 shows a scatterplot for errors corrected versus coverage including a linear regression model for 44 subjects (no time data for one student and two professionals), where the minimum time specified for a cell to be considered inspected/checked is >0.3 seconds. It is evident from Figure 2 that a moderate-strong relationship exists between the number of cells inspected and error detection/correction performance: $R^2$ value of 0.6421.

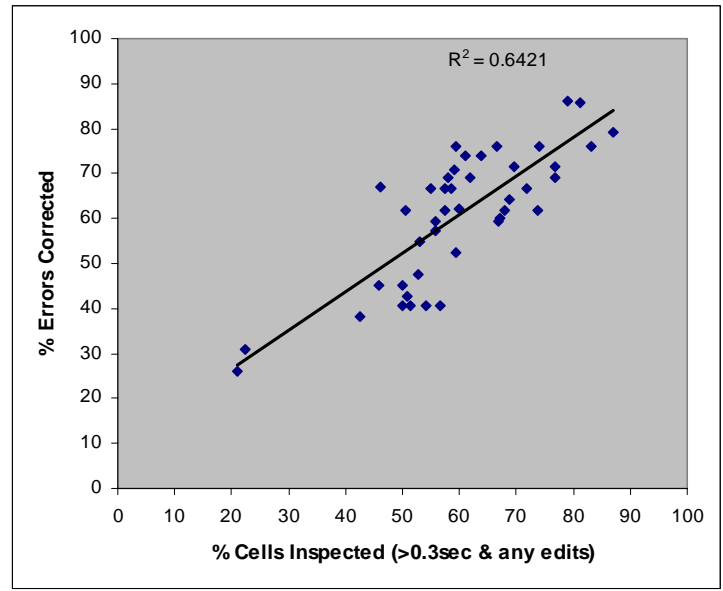

**Figure 2 – Errors Corrected over Cells Inspected (>0.3sec)**

Figure 3 shows a scatterplot for errors corrected versus coverage including a linear regression model for the same 44 subjects, with the same specified cell range, where the minimum time specified for a cell to be considered inspected/checked is >1 second: $R^2$ value of 0.607 shows a similarly moderate-strong linear relationship.

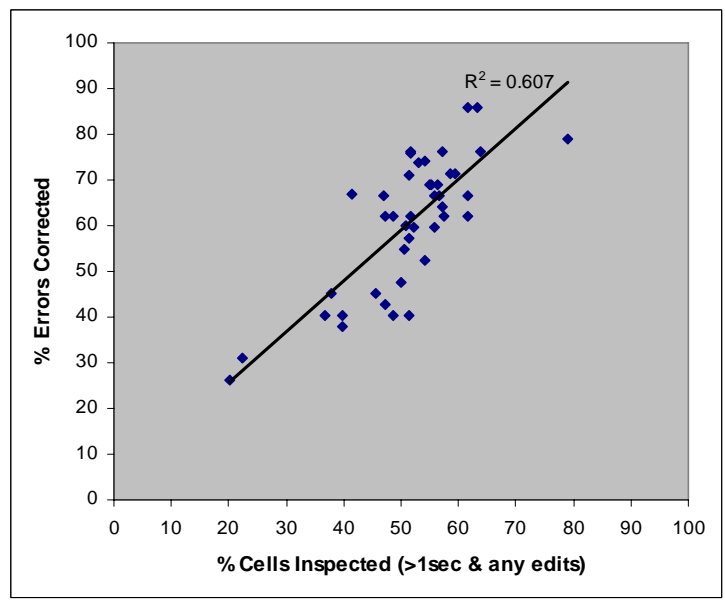

**Figure 3 – Errors Corrected over Cells Inspected (>1sec)**

A possible application of this research may be as a predictor of spreadsheet reliability based on relevant past error density and debugging performance information and the percentage of critical cells inspected.

The remainder of the analysis focuses on the debugging behaviour of professional participants.

### 4.3 Unique Formula Debugging Analysis: IF Statement

Initial analysis was carried out on the time based aspect of the spreadsheet trial regarding professional subjects debugging behaviour of a unique IF formula. The IF formula, see Appendix A – *Office Expenses F20*, was as follows:

=IF(F10+F18>7000,"Exceeds Limit","Within Limit")

The formula was incorrect, and should have been corrected by changing the 7,000 to 70,000. This change requirement was made known through instructions on the spreadsheet itself. Of the professional subjects, 38% amended the formula correctly. An aspect of professional subject's behaviour that stood out when the resulting data logs were analysed, is that nearly all professional subjects checked/debugged the formula exactly twice, with none amending the formula correctly on their second visit to the cell. The average times spent inspecting the cell for professional participants who amended the cell correctly was 14 seconds on the first visit and 12 seconds on the second visit, those who did not correct the cell formula spent slightly less time inspecting the cell. Although time based analysis has not been carried out on this formula cell for the student subjects, results showed that 58% of the students amended this cell correctly. It may be the case that the professional subjects easily understood the logic of the IF formula, and missed the simple error by not inspecting the formula more thoroughly.

### 4.4 Remote Sheet Reference Debugging

Analysis was carried out on the behaviour of professional participants in debugging an incorrect formula that included a remote cell reference (a reference to a cell value on a different worksheet within the same workbook). The original and correct value of this formula, see Appendix A – *Projections B19,* is as follows:

| Original Formula | Correct Formula |
|---|---|
| ='Office Expenses'!F10*4 | ='Office Expenses'!F18 |

Thirty-eight percent of professional participants corrected this formula error. It was identified that although only 38% of participants discovered and correctly amended the formula, 46% discovered the error but amended it incorrectly. This finding implies that while error detection and error correction are both part of the debugging process, they are still two distinct disciplines, and an individual's error detection capability should not be used as an automatic indication of their ability to correct errors found. Of the student subjects, only 12% corrected this error.

A reason for the very low correction rate for this formula could be that participants only discovered one of the two errors in the formula, and did not consider that there could be more than one error; a *mechanical* and a *logical* error. The cell value of *'Office Expenses'!F18* was a yearly estimate, calculated by a formula that added a range of 3-month estimates which had been multiplied by 4. An incorrect cell was referenced - a mechanical error, and multiplying the remote cell value by 4 was a logical error; as the relevant remote cell value had effectively been multiplied by 4 already. Of the 46% of professional subjects who discovered the error but amended it incorrectly, 66% of them found the mechanical error but missed the logical error.

Microsoft Excel provides a useful show-precedents auditing tool for inspecting cell formulas, where arrows indicate cells that are referred to by a formula. A problem with this tool occurs when a precedent is on another worksheet. Excel simply lets the user know that a remote cell is being referenced by displaying an icon. Double clicking this icon allows one to go directly to the referenced cell, which entails leaving the sheet currently being inspected. This can be very confusing. The error described above may have had a higher correction rate if relevant information on the remote cell was available without leaving the worksheet currently being inspected. This information could include the remote cells value and formula, the column and row headings associated with the remote cell, the remote cells precedents etc. Further analysis regarding the most and least detected errors was conducted, but is not detailed here.

## 5. CONCLUSION

Spreadsheets are designed, built and used by a variety of users, many of whom are not professional programmers and are not inclined towards following or learning software development and testing methodologies. This is a major contributing factor to the unreliability of spreadsheets. With the aim of aiding end-users in improving spreadsheet reliability, many spreadsheet auditing and debugging tools have been developed and made available, but tools should support users' natural debugging behaviour. This paper describes an experiment conducted with thirteen industry-based professionals and thirty-four accounting & finance students designed to unintrusively record end-user behaviour in spreadsheet error detection and correction activities. An experimental spreadsheet model was developed and subjects were asked to correct any errors found directly on the spreadsheet itself.

Overall results show that professionals (experts) are more efficient and effective spreadsheet debuggers than students (novices). Professional subjects outperformed student subjects in detecting and correcting errors of certain categories, namely *formula errors,* with a 16%-25% greater correction rate. Future analysis will aim at identifying the factors and behaviours that contribute to better debugging performance. An important finding is that a relationship exists between the percentage of critical cells inspected and the number of errors detected and corrected. In traditional software testing, predicting the reliability of software programs based on code coverage and defect density is a tried and tested method, which could possibly be applied to the spreadsheet paradigm. This study utilises a small, well-structured spreadsheet. But the question remains whether the findings can be applied to larger, poorly-structured spreadsheets. We believe that experts would outperform novices in debugging regardless, but that greater variance in debugging behaviour would occur with larger, real-world spreadsheets. The future aims of this study are to provide practical information for improving spreadsheet reliability by conducting further experiments and analysis in the near future, and possibly developing a spreadsheet debugging tool based on the experiment findings.

## Appendix A – Experimental Spreadsheet Model with Errors Colour Coded

### Error Colour Codes

| Error Types | |
|---|---|
| Clerical | |
| Rule Violation | |
| Data Entry | |
| Formula | |

### Payroll Worksheet

| | A | B | C | D | E | F | G | H | I |
|---|---|---|---|---|---|---|---|---|---|
| 1 | The spreadsheet has opened correctly, please proceed with the exercise | | | | | | | | |
| 2 | Choi Construction Company | | | | | | | | |
| 3 | | | | | Payroll for Week Ending: 15/03/20xx | | | | |
| 4 | | Code | Pay Rate | Regular Hours | Overtime Hours | Regular Pay | Overtime Pay | Total | Subtotals by Code |
| 5 | Adams | A | 8.9 | 40 | 3 | =C5*D5 | =C5*1.5*E5 | =F5+G5 | |
| 6 | Baker | A | 12.55 | 35 | 0 | =C6*D6 | 10 | =F6+G6 | |
| 7 | Carlton | A | 9.6 | 40 | 0 | =C7*D7 | 0 | =F7+G7 | =SUM(H5:H7) |
| 8 | Daniels | B | 10.2 | 35 | 2 | =C8*D8 | =C8*1.5*E8 | =F8+G8 | |
| 9 | Englebert | B | 9.6 | 40 | 5 | =C9D8 | =C9*1.5*E9 | =F9+G9 | |
| 10 | Franklin | B | 11.55 | 40 | 3 | =C10*D10 | =C10*1.5*E10 | =F10+G10 | =SUM(H7:H10) |
| 11 | Griffin | C | 10.8 | 400 | 2 | =C11*D12 | =1.5*C11*1.5*E11 | =F11+G11 | |
| 12 | Hartford | C | 9.9 | 40 | 10 | =C12*D12 | =C12*1.5*E12 | =F12+G12 | |
| 13 | Indio | C | 4.2 | 40 | 0 | =C13*D13 | =C13*1.5*E13 | =F13+G13 | |
| 14 | Jackson | C | 21.5 | 40 | 11 | =C14*D14 | =C14*1.5*E14 | =F14+G14 | =SUM(H10:H14) |
| 15 | | | | | | | | | |
| 16 | Totals | | | =SUM(D5:D14) | =SUM(E5:E14) | =SUM(F5:F14) | =SUM(G5:G13) | =SUM(H6:H13) | =H16 |
| 17 | | | | | | | | | |

Payroll / Office Expenses / Projections /

### Office Expenses Worksheet

| | A | B | C | D | E | F |
|---|---|---|---|---|---|---|
| 1 | Choi Contruction Company | | | | | |
| 3 | | | | Office Expenses for First Three Months of 20xx and Estimated for Year | | |
| 4 | Variable Expenses | Jan | Feb | Mar | Total | Year (Est) |
| 5 | Electricity | 245 | 261 | 221 | =SUM(B5:D5) | =3*E5 |
| 6 | Telephone | 1350 | 2350 | 1175 | =SUM(B6:D6) | =4*E6 |
| 7 | Heating (Gas) | 383 | 456 | 403 | =SUM(B7:D7) | =4*E7 |
| 8 | Subscriptions | 117 | 113 | 150 | =3*D8 | =4*E8 |
| 9 | Petty Cash | 250 | 275 | 290 | =SUM(B9:D9) | =4*E9 |
| 10 | Total: | | | | | =SUM(F5:G9) |
| 12 | Fixed Expenses | Jan | Feb | Mac | Total | Year (Est) |
| 13 | Rent | 2500 | 2500 | 2500 | =SUM(B13:D13) | =4*E13 |
| 14 | Water | 250 | =C13/10 | 250 | =SUM(B14:D14) | =4*E14 |
| 15 | Custodial(Cleaning) | 130 | 130 | 130 | =SUM(B15:D15) | =4*E15 |
| 16 | Maintenance Fees | 350 | 350 | 330 | =SUM(B16:D16) | =4*E16 |
| 17 | Food Services | 450 | 450 | -450 | =SUM(B17:D17) | =4*E17 |
| 18 | Total: | | | | | =SUM(F13:F16) |
| 19 | | | | | | |
| 20 | Variable + Fixed Expenses exceed 70,000 limit? | | | | | =IF(F10+F18>7000,"Exceeds Limit","Within Limit") |
| 21 | | | | | | |
| 22 | | | | | | |

Payroll / Office Expenses / Projections /

**Projections Worksheet**

| | A | B | C | D | E | F | G |
|---|---|---|---|---|---|---|---|
| 1 | **Choi Construction Company** | | | | | | |
| 2 | Today's Date | 04/05/2006 | | | | | |
| 3 | | | | | | | |
| 4 | | | | | | | |
| 5 | Projection Rates: | | | | | | |
| 6 | Payroll A Employees | 0.045 | | | | | |
| 7 | Payroll B Employees | 0.049 | | | | | |
| 8 | Payroll C Employees | 0.042 | | | | | |
| 9 | Office Variable Expenses | 0.023 | | | | | |
| 10 | Office Fixed Expenses | 0.035 | | | | | |
| 11 | | | | | | | |
| 12 | | | | | | | |
| 13 | | Current Year | Year 1 | Year 2 | Year 3 | Year 4 | Year 5 |
| 14 | **Payroll Expenses** | | | | | | |
| 15 | Category A | =52*Payroll!7 | =B15*(1+$B$6) | =C15*(1+$B$6) | =D15*(1+$B$6) | =E15*(1+$B$6) | =F15*(1+$B$6) |
| 16 | Category B | =52*Payroll!10 | =B16*(1+$B$7) | =C16*(1+$B$7) | =D16*(1+$B$7) | =E16*(1+$B$7) | =F16*(1+$B$7) |
| 17 | Category C | =50*Payroll!16 | =B17*(1+$B$8) | =C17*(1+$B$8) | =D17*(1+$B$8) | =E17*(1+$B$8) | 235600 |
| 18 | **Office Expenses** | | | | | | |
| 19 | Fixed Expenses | =Office Expenses!F10*4 | =B19*(1+$B$9) | =C19*(1+$B$9) | =D19*(1+$B$9) | =E19*(1+$B$9) | =F19*(1+$B$9) |
| 20 | Variable Expenses | =Office Expenses!F8*4 | =B20*(1+$B$9) | =C20*(1+$B$9) | =D20*(1+$B$9) | =E20*(1+$B$9) | =F20*(1+$B$9) |
| 21 | | | | | | | |
| 22 | Total | =SUM(B15:B20) | =SUM(C15:C20) | =SUM(D15:D20) | =SUM(E15:E20) | =SUM(F15:F20) | =G15+G16+G17+G19+G20+G21 |
| 23 | | | | | | | |
| 24 | Does yearly increase in expenses exceed 4% | =IF(C22/B22>1.04,"Yes","No") | =IF(D22/C22>1.04,"Yes","No") | =IF(E22/D22>1.04,"Yes","No") | =IF(F22/E22>1.04,"Yes","No") | =IF(G22/F22>1.04,"Yes","No") | |
| 25 | | | | | | | |

Payroll / Office Expenses \ **Projections** /